\documentclass[twocolumn]{aa}
\usepackage[usenames,dvipsnames]{color}
\usepackage{gensymb}
\usepackage{graphicx}
\usepackage{amsmath, amssymb}
\usepackage{lineno}

\usepackage{txfonts}
\begin{document} 

   \title{Revised historical solar irradiance forcing}

   \author{T. Egorova
                      \inst{1},
                       W. Schmutz 
                     \inst{1},  
                      E. Rozanov
                     \inst{1,2},
                      A. I. Shapiro
                     \inst{3},
                      I. Usoskin
                     \inst{4},
                     J. Beer
                     \inst{5},
                     R. V. Tagirov
                     \inst{6},
                      \and
                      T. Peter
                     \inst{2}
                             }

   \institute{Physikalisch-Meteorologisches Observatorium Davos, World Radiation Center, Davos Dorf, Switzerland
   \and
   Institute for Atmospheric and Climate Science, ETH, Zurich, Switzerland
   \and
   Max-Planck-Institut f\"{u}r Sonnensystemforschung, G\"{o}ttingen, Germany
   \and 
   Space Climate Research Unit and Sodankyl\"{a} Geophysical Observatory, University of Oulu, Finland
   \and
   EAWAG, D\"{u}bendorf, Switzerland
   \and
   Imperial College London, Astrophysics Group, Blackett Laboratory, London, United Kingdom
   }


 
  \abstract
   {There is no consensus on the amplitude of the historical solar forcing. The estimated magnitude of the total solar irradiance difference between Maunder minimum and present time ranges from 0.1 to 6\,W/m$^2$ making uncertain the simulation of the past and future climate. One reason for this disagreement is the applied evolution of the quiet Sun brightness in the solar irradiance reconstruction models. This work addresses the role of the quiet Sun model choice and updated solar magnetic activity proxies on the solar forcing reconstruction.}
   {We aim to establish a plausible range of the solar irradiance variability on decadal to millennial time scales.}
   {The spectral solar irradiance (SSI) is calculated as a weighted sum of the contributions from sunspot umbra/penumbra, faculae and quiet Sun, which are pre-calculated with the spectral synthesis code NESSY. We introduce activity belts of the contributions from sunspots and faculae  and a new structure model for the quietest state of the Sun. We assume that the brightness of the quiet Sun varies in time proportionally to the secular (22-year smoothed) variation of the solar modulation potential.}
   {A new reconstruction of the TSI and SSI covering the period 6000 BCE - 2015 CE is presented. The model simulates solar irradiance variability during the satellite era well. The TSI change between the Maunder and recent minima ranges between 3.7 and 4.5\,W/m$^2$ depending on the applied solar modulation potential. The implementation of a new quietest Sun model reduces, by approximately a factor of two, the relative solar forcing compared to the largest previous estimation, while the application of updated solar modulation potential increases the forcing difference between Maunder minimum and the present by 25-40\,\%.}
 {}
   \titlerunning{Reconstruction of historical solar forcing}
   \authorrunning{Egorova et al.}
   \maketitle
%
\section{Introduction}

Climate change prediction using sophisticated numerical models with a sufficient degree of accuracy is one of the most important issues in modern climate science. Despite the impressive progress in climate modelling during the last 10-15 years, many problems with climate models still persist \citep{Flato2013} because simulation results are largely determined by the set of the input parameters and boundary conditions which are known to within uncertainties. One of the most ambiguous and least understood is the solar forcing \citep{Myhre2013}, which is important to quantify to understand the influence of natural forcing on Earth's climate. 

Several reconstructions of the historical solar irradiance variability using different approaches have been performed \citep[e.g.,][]{Wang2005, Tapping2007, Steinhilber2009, Krivova2010, Vieira2011, Shapiro2011} suggesting changes in TSI between Maunder minimum and present day from 0.1\,W/m$^2$ to 6\,W/m$^2$, which reveals no consensus on the amplitude of the solar forcing \citep{Solanki2013, Kopp2016}. One of the largest increase of the total solar irradiance (TSI) of 6$\pm$3\,W/m$^2$ between the Maunder minimum (1645-1715) and the Modern Grand Solar maximum was suggested by \citet[hereafter SSR11]{Shapiro2011}, implying a potentially large effect on climate. 

The results of the solar irradiance reconstruction from SSR11 were used to simulate a climate response to such a centennial scale solar irradiance variation \citep[e.g.,][]{Anet2014, Feulner2011, Schurer2013}. However, the results did not provide a definite conclusion about the reliability of the large solar forcing as suggested by SSR11. While \cite{Feulner2011} and \cite{Schurer2013} concluded that a large solar forcing is not consistent with the observed climate change, \citet{Anet2014} demonstrated that a large solar forcing is necessary to properly reproduce colder climate during the Dalton minimum of the solar activity (1790-1830 CE).      

The SSR11 approach was re-evaluated by \citet{Judge2012} who noticed that the solar model for the minimum state of the quiet Sun is too cold and recommended some revision. The calculations of the 11-year SSI variability in SSR11 were kept simple and several problems with this approach have also been discussed in the literature \citep{Ermolli2013, Shapiro2013, Yeo2014}. 

In this paper we address these shortcomings, of both the 11-year and longer time scales, by improving the SSR11 approach to provide a more reliable reconstruction of solar irradiance variability on decadal to millennial time scales for the climate community. The main goal of the paper is to present and discuss a new spectral (from 120 to 100000 nm) and total solar irradiance time series since 6000 BCE using an updated solar irradiance model reconstruction, and solar activity proxies such as sunspot number and solar modulation potential retrived from the time series of cosmogenic radionuclides in ice cores.    

\section{Method}

To obtain secular changes of the spectral solar irradiance on time scales from years to millennia, we use the model CHRONOS (Code for the High spectral ResolutiOn recoNstructiOn of Solar irradiance), which calculates the coverage of the solar surface by different solar structures (sunspots, faculae and quiet Sun) and integrates their spectra. CHRONOS is a new version of the model used in SSR11, with an updated treatment of the filling factors and long term variability of the quiet Sun irradiance. The improvements implemented allow for an increase in the efficiency of the code and in the accuracy of the solar irradiance calculations. 

CHRONOS is based on the new spectral synthesis code NESSY (NLTE Spectral SYnthesis) \citep{Tagirov2017}. We consider four solar surface components: quiet Sun, faculae, sunspot umbra, and sunspot penumbra. Thus, the time evolution of SSI at any moment in time $t$ is represented by

\begin{equation}
  \begin{split}
F(\lambda,t) = F_{qs}(\lambda,t) +\alpha_f(t)(F_f(\lambda)- F_{qs}(\lambda,t)) \\ 
+ \alpha_{su}(t)(F_{su}(\lambda) - F_{qs}(\lambda,t)) \\
+\alpha_{sp}(t)(F_{sp}(\lambda)- F_{qs}(\lambda,t)),
  \end{split}
     \label{Eq1}
\end{equation}

\noindent
where $F(\lambda,t)$ are the full disk SSIs calculated with NESSY using solar atmospheric structures representing: 
the time variable quiet Sun spectrum $F_{qs}(\lambda,t)$,
faculae $F_f(\lambda)$,
sunspot umbra $F_{su}(\lambda)$,
and sunspot penumbra $F_{sp}(\lambda)$.
The coefficients
$\alpha_i(t)$ are the filling factors, representing the projected fractional area of the solar disk covered by these magnetic features at the time $t$, relative to the total surface of the Sun. 
Our approach differs from other published reconstructions by the inclusion of significant variation of the quiet Sun irradiance, as described below.

\subsection{Computing the long-term evolution of the quiet Sun irradiance $F_{qs}(\lambda,t)$}

\begin{figure}
   \centering
   \includegraphics[width=9cm]{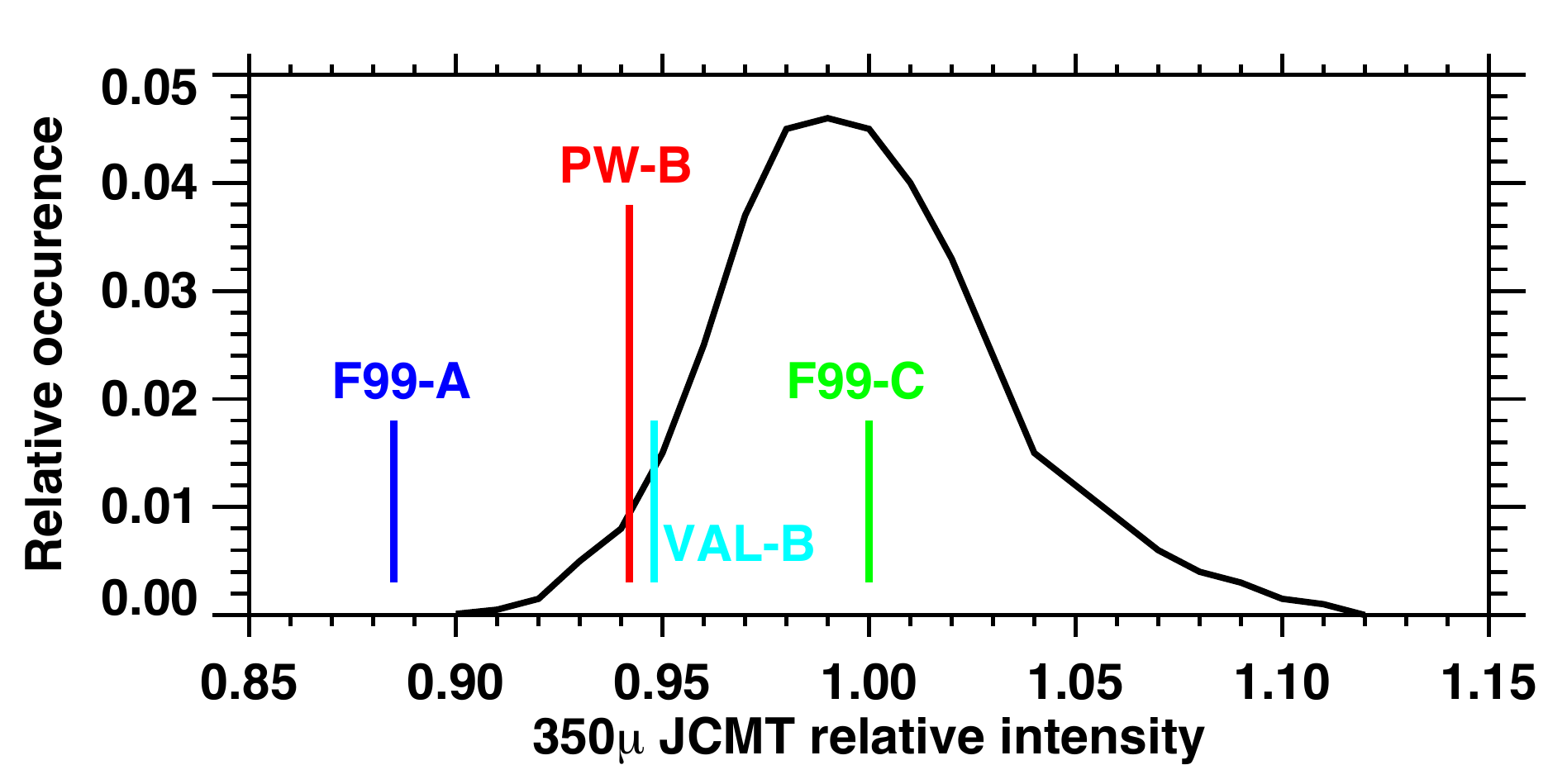}
   \caption{Distribution of the relative brightness of quiet Sun regions at  350\,$\mu$m observed by the James Clerk Maxwell telescope by
 \citet{Lindsey1995} and analyzed by \citet{Judge2012}. The radiance emitted at 350\,$\mu$m originates close to the temperature minimum of the solar atmosphere structure. The relative brightness 1.00 corresponds to the median of the frequency distribution which is set equal to the radiance of the model F99-C at  350\,$\mu$m, as this model is reproducing successfully the present day quiet Sun \citep{Shapiro2010}. The radiance calculated with the model F99-A, model VAL-B and the model PW-B, are indicated relative to the model F99-C. The relative brightness of the model VAL-B is placed as given by \citet{Judge2012}, whereas the the other indicated locations have been obtained from the relative radiance obtained from NLTE computations with NESSY.}
   \label{Fig1}
   \end{figure}
   
   The long-term evolution of the quiet Sun irradiance $F_{qs}(\lambda,t)$  in Eq. \ref{Eq1} is assumed to be driven by the slowly changing content of the small-scale magnetic fields. Figure \ref{Fig1} illustrates the brightness distribution as measured at 350\,$\mu$m  by the James Clerk Maxwell telescope \citep{Lindsey1995} and analysed by Judge et al. (2012). The brighter regions differ from the fainter ones in that they contain more small-scale active regions with magnetic fields too weak to be detected with present-day instrumentation. There are different approaches to account for these fields, and these are the reason for the uncertainties. The question, on which presently there is no observational answer, is how the quiet Sun brightness is reduced when it is in a grand minimum as e.g. in the Maunder minimum. SSR11 assumed that this distribution at minimal activity is represented best by the faint super-granule cell interior because this is a region on the Sun with the least amount of magnetic field. The representation of this region by the atmospheric solar model chosen by SSR11 turned out to be incorrect as outlined below. However, we still consider and adapt the physical assumptions of SSR11 as the basis for our study. It was assumed by SSR11 that the contribution to the overall radiated energy by small scale magnetic activity can be described with a proxy, i.e. using the 22-year mean of the solar modulation potential, $\phi(t)$. This 22-year average also includes the contribution of two 11-year magnetic activity cycles, which in principle is already accounted for by the other terms in Eq. \ref{Eq1}. A more appropriate proxy would be the lower envelope of the yearly modulation potential since the minima correspond to the quiet Sun. However, the yearly $\phi$-values are only available for the recent past and therefore, we use the mean instead of the lower envelope as a proxy. An investigation of yearly values reveals that a lower envelope and a mean are almost linearly proportional to each other. As we are using the $\phi$-proxy only in a relative way, the mean proxy value realisation is almost identical (within 10\%) to the lower envelope. However, a mixed treatment of using a lower envelope from an yearly dataset (if available), together with a 22-year mean would lead to inconsistencies: we have to make sure that equal activity levels in the past and present agree and yield the same quiet Sun irradiance.  Accordingly, changes of the quiet Sun irradiance are described by the following equation
   
\begin{equation}
F_{qs}(\lambda,t)=F_C(\lambda)+(F_B(\lambda)-F_C(\lambda)) \frac{\phi_{ref}-\phi(t)}{\phi_{ref}-\phi_{min}}\, ,
     \label{Eq2}
  \end{equation} 
\noindent
where $\phi(t)$ is the time evolving 22-year-mean solar modulation potential for a given time $t$.  $\phi_{ref}$ is the solar modulation potential, which is evaluated as the mean of the period 1974-1996. $\phi_{min}$ represents the smallest solar modulation potential during the time period considered (6000 BCE - 2015),
which in all cases is close to the year 1470 (the deep phase of the Sp\"{o}rer minimum). $F_B(\lambda)$ is the SSI resulting from the present work solar model B (PW-B, see below), representing the minimum solar activity level, and $F_C(\lambda)$ is the SSI of the present day quiet Sun. The relative use of $\phi$ eliminates uncertainties related to the absolute level of the solar modulation potential \citep{Usoskin2005,Herbst2010}. 
   
Following SSR11 we assume that the irradiance of the quiet Sun is varying in time due to the brightness of small-scale magnetic fields, which in turn depend on the level of solar magnetic activity in the preceding decades. This hypothesis is the key assumption of our approach. For a more detailed discussion of this assumption, the reader is referred to SSR11. 

The basic assumption behind Eq. \ref{Eq2} is that the irradiance of the quiet Sun $F_{qs}(\lambda,t)$ varies between two boundaries given by the modern maximum $F_C(\lambda)$ which is known from observations and treated as reference and the expected lowest irradiance of the quiet Sun $F_B(\lambda)$ attributed to the lowest magnetic activity derived from $\phi$. $F_{qs}(\lambda,t)$ is then linearly scaled between these two boundaries by $\phi$ for any time $t$. Most critical is the choice of $F_B(\lambda)$, which determines the amplitude of the changes.

The range of the quiet Sun irradiance variability is crucial for the resulting amplitude, but is arbitrary to some degree because modern experiments have not measured the properties of solar irradiance outside the present period of high solar activity. SSR11 argued for the variation range to be between the A and C solar models by Fontenla et al.(1999), hereafter termed F99. Their reason for choosing model A for the quietest state was to select an existing quiet Sun model, which should basically represent the faint 1/8$^{th}$ quantile of the statistical brightness distribution of the quiet Sun. It was then pointed out by \citet{Judge2012} that the model F99-A produces a flux at 350 $\mu$m outside the observed brightness distribution, which disqualifies this model for the intended use.  Figure \ref{Fig1} illustrates by how much the model F99-A lies outside the observations. \citet{Judge2012} recommended to use instead a model that lies close to the relative brightness indicated by VAL-B in Figure \ref{Fig1}, which is a model from the set representing atmospheric structures by \citet{Vernazza1981}. 
 
We introduce here a new solar atmosphere structure, which is constructed such that it represents what SSR11 originally intended: the quiet Sun during extremely low magnetic activity should be represented by a model at the faint 1/8$^{th}$ quantile of the statistical brightness distribution of the quiet Sun. The model sets published by \citet{Vernazza1981} and \citet{Fontenla1999} differ in temperature and density structure at large optical depth. This should not be the case if the models are used for irradiance contrast as in Eq. \ref{Eq1}, because these models should converge to the same structure at large depths (except for the sunspot models, for which the structures are strongly influenced by a concentrated magnetic field). Therefore, we should not use VAL-B; instead a model that fits to the F99-set is needed.

\citet{Fontenla1999} did not publish a structure that is between their A and C structures, but it is straightforward to interpolate between their models. We calculate the mean of the temperatures of the models A and C for a given height and then calculate the density by ensuring pressure equilibrium between all three model structures. We term this new solar atmosphere structure as present work model B (PW-B) and the resulting relative brightness of this atmosphere model at 350\,$\mu$m is indicated in Figure \ref{Fig1}. Given the limited knowledge available on the possible minimal state the Sun could reach, its relative intensity to the F99-C model is in full agreement with the recommended model property to represent a minimal-activity state of the quiet Sun.

If we evaluate Eq. \ref{Eq2}  for the past two solar minima in 1996 and 2009, the relative $\phi$-difference term yields about 4\,\%\ of the $F_B(\lambda)-F_C(\lambda)$ forcing difference, which translates to about 0.2\,W/m$^2$ decrease in TSI between the two minima according to our approach. This value is still below the threshold of present TSI composite uncertainties \citep{Ball2016a} and therefore, it is not yet possible to validate observationally the proposed relation of Eq.\ref{Eq2}. 

For the calculation of SSI we use several solar modulation potential datasets that have been recently published: McCracken and Beer (2017, personal communication), referred hereafter as PHI-MC17; \citet[]{Usoskin2016} referred hereafter as PHI-US16, and \citet[]{Muscheler2016} referred as PHI-MU16. The use of several independent $\phi$ time series allows for a better estimate of the uncertainty range of the reconstructed SSI.

\subsection{Computation of the short-term variations of the spectral solar irradiance}

The short-term irradiance reconstruction follows the approach by \citet{Krivova2003}, who describe SSI at any particular wavelength as the sum of several individual components. This model distinguishes several types of solar structures. However, not all of them can be considered here because detailed information about the distribution and strength of the solar magnetic features is not always available, especially for the historic reconstructions and future projections.

For a simplification we assume that the contribution of the active regions is calculated for present day conditions and remain constant in time. Therefore, the time evolving contribution from sunspots and faculae comes solely from the time varying filling factor $\alpha_i(t)$. This assumption completely holds for the satellite era, but cannot be strictly proven for the long-term due to an absence of observations. Filling factors are calculated directly from the sunspot number as described in section\,2.3 instead of the more detailed approach of \citet{Krivova2003}, which is based on high resolution solar imagery. This simplification is necessary because no detailed information on the solar magnetic field is available for an historic reconstruction. The actual evolution of the sunspot number (SSN) for 1900-2016 was taken from version 2 of the WDC-SILSO data set prepared by the Royal Observatory of Belgium, Brussels (http://www.sidc.be/silso/datafiles). For the 1749-1899 and 1612-1748 we applied SSN calculated from sunspot group number (SGN) time series published respectively by \cite{Chatzistergos2017}  and \cite{Lockwood2014}. For the years before 1612 the decadal scale SSN variability is represented by sine function with the magnitude proportional to the solar modulation potential. 

The irradiance spectra are computed based on model atmosphere structures adopted from F99. The model F99-C represents the present day quiet Sun, model F99-P represents an average faculae region, and model F99-S describes an average sunspot umbra. The model set of F99 does not contain a model for the penumbral region of a sunspot. Since we only need the penumbral contrast relative to the quiet Sun (and not the penumbra spectrum itself) we followed the approach of \citet{Shapiro2015} and employed penumbra and quiet Sun models from \citet{Fontenla2006} to calculate the penumbral contrast as a function of wavelength and disk position.

\subsection{Activity belts}

In calculating the contribution of sunspots and faculae to solar irradiance variability, SSR11 assumed the uniform coverage of the solar disk by these features (hereafter, the full disk distribution). However, spots and faculae mainly appear in the so-called activity belts, i.e. at latitudes  between approximately $\pm 5^{\circ}$ and  $\pm 30^{\circ}$ for spots and between approximately $\pm 5^{\circ}$ and  $\pm 40^{\circ}$ for faculae \citep[see e.g.][]{Wenzler2006, Krivova2007}. In the present study we have accounted for such a non-uniform distribution of active regions and outline below our new implementation and its effect on the irradiance reconstructions.

The contribution of any active region (i.e. spot or faculae) to solar irradiance can be written in the following form:

\begin{equation}
\Delta F_{i} (t,\lambda)= \iint\limits_{\mathit{solar \,\, disc}}  \left ( I_{i} \left ( \vec{\lambda, r}  \right )    -  I_{qs} \left ( \vec{\lambda, r}  \right )  \right ) \, {\cal A}_{i} (t, \vec{r}) \, \, d \Omega,
\label{eq:irr}
\end{equation}
where $  I_{i} \left ( \vec{\lambda, r}  \right )  $ and $ I_{qs} \left ( \vec{\lambda, r} \right )$ are solar irradiance intensities at the wavelength $\lambda$ along the direction $\vec{r}$ from the active region and quiet Sun, respectively. The time-dependent function ${\cal A}_{i} (t, \vec{r}) $ represents the fractional coverage of the solar disk by active regions along the direction $\vec{r}$ and $d \Omega$ is a solid angle. The function  ${\cal A}_{i} (t, \vec{r}) $ and the filling factors $\alpha_{i}(t)$ introduced in Sect.~2 are determined from observations (see Sect.~{2.2})  are connected via

\begin{equation}
\alpha_{i}(t)=\iint\limits_{\mathit{solar \,\, disc}} {\cal A}_{i} (t, \vec{r}) \, d \Omega/\Omega_{\odot},
\label{eq:ff}
\end{equation}
where $ \Omega_{\odot}  $  is the solid angle subtended by the solar disk. In case of the full disk distribution of active features Eq.~(\ref{eq:ff}) results in ${\cal A}_{i} (t, \vec{r})  \equiv \alpha_{i}(t)$. In case of the belt distribution  ${\cal A}_{i} (t, \vec{r}) $ is equal 
$\alpha_{i}(t) \,  \Omega_{\odot}  /  \Omega_{\rm belt} $ (where $\Omega_{\rm belt}$  is solid angle subtended by the considered activity belts) within the activity belt and to zero outside of the activity belts.

If the contrast of active features  $  \left ( I_{i} \left ( \vec{\lambda, r}  \right )    -  I_{qs} \left ( \vec{\lambda, r}  \right )  \right )$  was independent of the position on the solar disk then it could be taken out of the integration in Eq.~(\ref{eq:irr}). Consequently, the contribution of active regions to solar irradiance would have been independent of the assumed distribution of the active features. In reality, the contrast is a strong function of the positions on the solar disk due to the centre-to-limb variation (CLV).

\begin{figure}
   \centering
   \includegraphics[width=9cm]{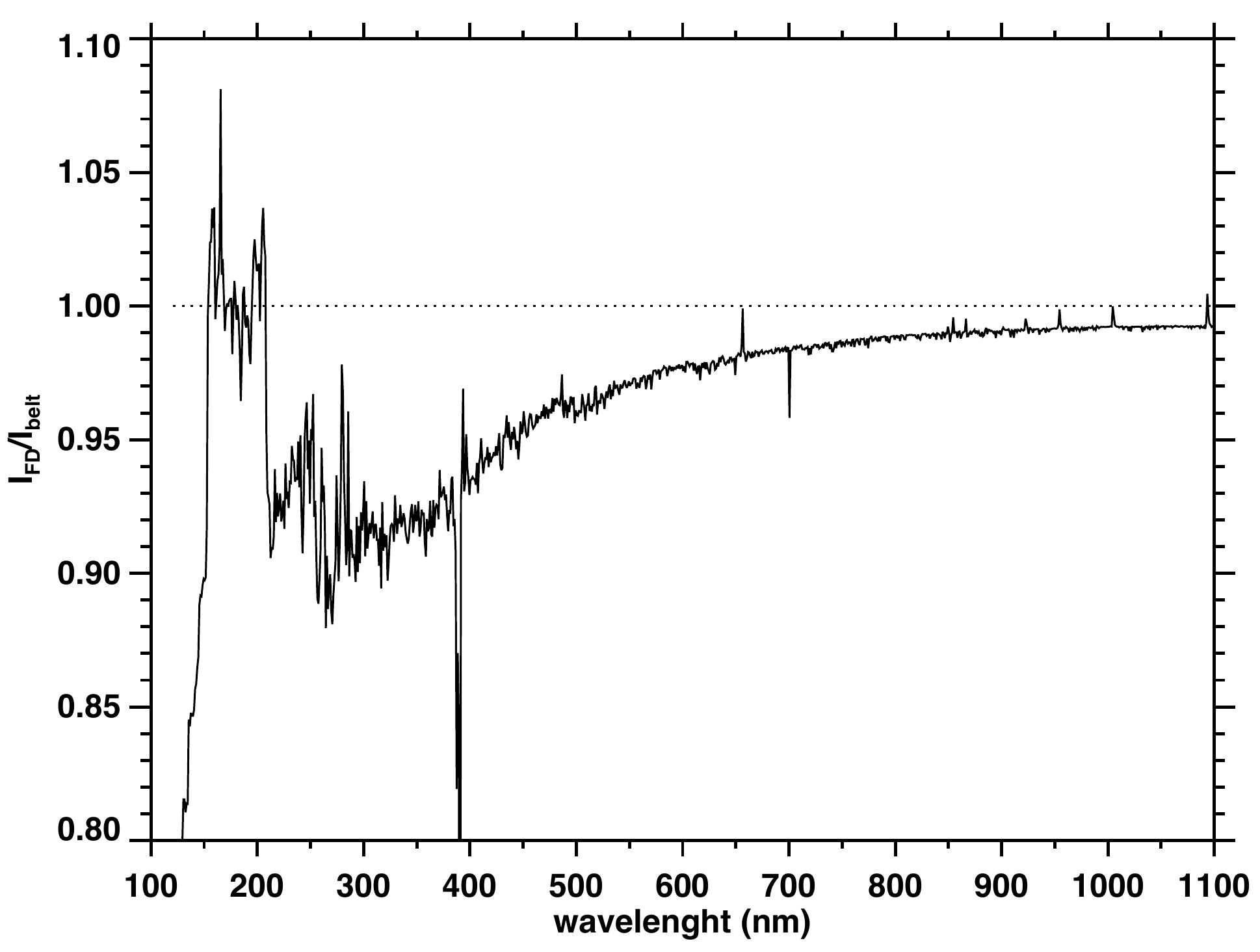}
   \caption{Ratio of the quiet Sun irradiance from the full solar disk to that from the facular activity belt scaled by $\Omega_{\odot}  /  \Omega_{\rm belt} $  to take into account the difference in the surface area.}
   \label{Fig2}
   \end{figure}

To illustrate the effect of the CLV we plot in Figure \ref{Fig2} the ratio between emergent quiet Sun intensities averaged over the full disk ($I_{\rm FD} =  \int\limits_{\mathit{solar \,\, disc}}  I_{qs} \left ( \vec{\lambda, r}  \right ) d \Omega  / \Omega_{\odot}$) and over the facular belt  ($I_{\rm belt} =\int\limits_{\mathit{belt}}  I_{qs} \left ( \vec{\lambda, r}  \right ) d \Omega  / \Omega_{\rm belt} $). In the case of limb darkening the ratio is below unity, while in the case of the limb brightening the ratio is greater than unity, because the contribution from the limb is smaller for the belts than for the full disk.

Figure \ref{Fig2} demonstrates that the limb darkening dominates for all wavelengths except 150-200\,nm. The CLV effect is 
most important below 150\,nm and between 200 and about 500\,nm. At longer wavelengths the Planck function becomes less sensitive to the temperature change so that the effect is small, even though the limb and disk center radiation is formed in regions with quite different temperatures. Strong photospheric and low chromospheric spectral lines shift the formation height to the temperature minimum region and thus decrease the CLV effect. This is observed in Figure \ref{Fig2}  as strong upward jumps, bringing the ratio almost to unity in the infrared. Spectral lines of moderate strength have the opposite effect, increasing the CLV and this appears as numerous weak downward jumps. The reason for this is that the moderate and weak lines have a stronger effect on the limb radiation than on the disk center.

   \begin{figure}
   \centering
   \includegraphics[width=9cm]{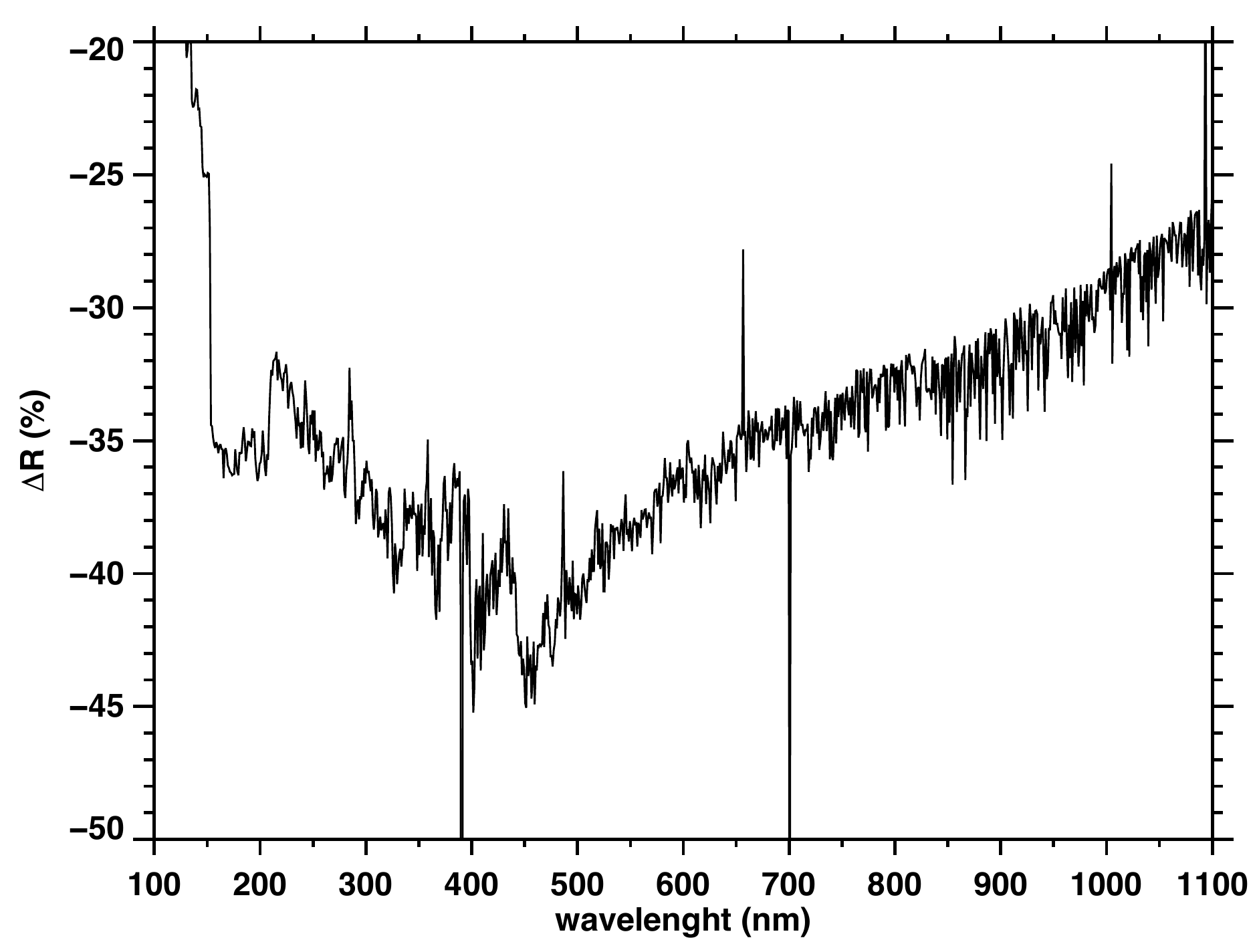}
   \caption{Relative deviation of the quiet Sun irradiance from the full solar disk to that from the facular activity belt scaled by $\Omega_{\odot}  /  \Omega_{\rm belt} $ to take into account the difference in the surface area.}
   \label{Fig3}
   \end{figure}
   
Figure  \ref{Fig3} shows the relative deviation of the facular contribution to the brightening of the Sun calculated using the activity belt approach compared to that using 
the full disk model. It demonstrates that application of the full disk model overestimates the facular contribution by 35-45\,\% from 200 to 450 nm. For the longer wavelengths the overestimation gradually decreases from 45 to 25\% at 1100 nm.   
\subsection{Sunspot and facular disk area coverages as functions of sunspot number}

In order to estimate filling factors for active regions for the pre-satellite epochs we derive a relationship between the filling factors and sunspot number for the period when both magnetograms and sunspot observations are available. Daily data of sunspot and facular filling factors have been taken from \citet{Ball2014}, who based their filling factors on: (1) full-disk magnetograms and continuum images from KP/512 (National Solar Observatory Kitt Peak Vacuum Tower, 512-channel diode array) for the period from 23.08.1974 to 04.04.1992 (1734 days); (2) KP/SPM (Kitt Peak spectromagnetograph) for the period  from 21.11.1992 to 21.09.2003 (2055 days); and (3) SOHO/MDI (Michelson Doppler Imager) from 19.02.1999 to 02.10.2010 (3799 days). Daily solar spot numbers (SSN) were taken from version 2 of the SILSO dataset prepared by the Royal Observatory of Belgium, for the overlapping period from 23.08.1974 to 02.01.2010 (13190 days). For the period from 19.02.1999 to 21.09.2003, when data from both KP/SPM and SOHO/MDI are available, we used the latter.
 
\begin{figure}
   \centering
   \includegraphics[width=9cm]{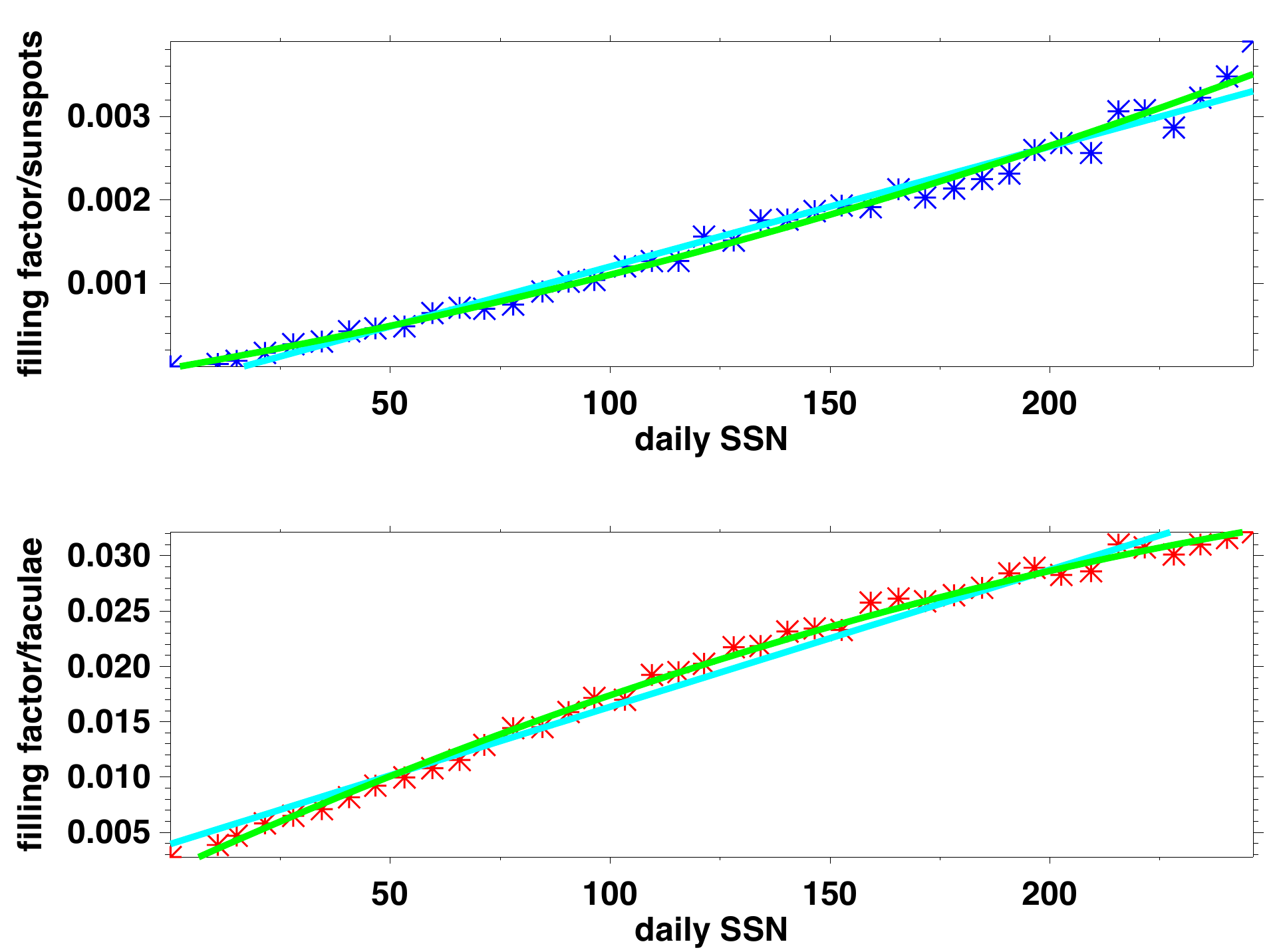}
   \caption{Daily sunspot (top, blue asterisks) and facular (bottom, red asterisks) filling factors grouped into 40 bins as a function of SSN. Light blue line represents linear and green line represents second order polynomial fits.}
  \label{Fig4}
   \end{figure}

For the regression analysis we binned the filling factors into 40 equal size bins covering the SSN range from 0 to 250 and applied a linear fit to sunspot and facular filling factors as a function of the SSN. For facular filling factors, we applied both a linear and second-order polynomial fit. Figure  \ref{Fig4} shows the results of the regression analysis and Tables \ref{Table1} and \ref{Table2} list the regression coefficients, their 1-sigma uncertainty range, and the RMS (root mean square) difference to the linear and second order polynomial fits. Figure \ref{Fig4} illustrates that sunspot filling factors follow an linear relationship to SSN, whereas  faculae filling factors are nonlinear. For the calculation of the filling factors for sunspot umbra and penumbra we used the ratio $\alpha_{su}$:$\alpha_{sp}$ = 1:4 (Wenzler et al., 2006).

\begin{table}
\caption{Linear fit to the disc area coverage.}  
\centering     
\label{Table1}      
\footnotesize
\begin{tabular}{c c c} 
\hline       
{Coefficients} &{Sunspot} & {Faculae} \\ 
\hline                    
a &{$(-24 \pm 5) \times 10^{-5}$} & {$(40 \pm 4) \times 10^{-4}$}   \\ 
\hline
b &{$(144 \pm 3) \times 10^{-7}$} & {$(124 \pm 3) \times 10^{-6}$}   \\ 
\hline	
RMS & {$5.3 \times 10^{-7}$} & {$5.9 \times 10^{-5}$} \\
\hline 
\end{tabular}                 
\tablefoot{
Listed are the regression coefficients, their 1-$\sigma$ uncertainty range and the RMS difference for the linear fit: y = a + b*SSN, where y is the filling factor and SSN the Sunspot number.}
\end{table}

\begin{table}
\caption[ ]{Quadratic fit to the disc area coverage.} 
\centering
\label{Table2}      
\footnotesize
\begin{tabular}{c c c} 
\hline       
{Coefficients} &{Sunspot} & {Faculae}\\ 
\hline                    
a &{$(-1.6 \pm 6) \times 10^{-5}$} & {$(16.1 \pm 3) \times 10^{-4}$}   \\ 
\hline
b &{$(9.1 \pm 1) \times 10^{-6}$} & {$(180 \pm 5) \times 10^{-6}$}   \\ 
\hline
c &{$(21 \pm 4) \times 10^{-9}$} & {$(-23 \pm 2) \times 10^{-8}$}   \\ 
\hline
RMS & {$9.3 \times 10^{-7}$} & {$1.4 \times 10^{-5}$} \\ 	
\hline                  
\end{tabular}
\tablefoot{
Listed are the regression coefficients,  their 1-$\sigma$ uncertainty range and chi-squared statistic for polynomial fit: y = a + b*SSN + c*SSN$^2$ , where y is the filling factor. and SSN the Sunspot number.}
\end{table}  

\subsection{The Ly\,$\alpha$ line}
$Ly\,\alpha$ line is important for the proper simulation of terrestrial atmospheric temperature and chemistry. \citet{Schoell2016} analysed the results of the radiative transfer code COSI and reported that both the absolute value and variability of the reconstructed $Ly\,\alpha$ flux are strongly underestimated. The disagreement with the absolute value has been resolved by using NESSY, the new version of the radiative transfer code, which is able to properly simulate solar irradiance below 130 nm (see \citet[]{Tagirov2017} for details).  We calculated an annual mean irradiance of 6.1 mW/m$^2$ for the year 2008 for the $Ly\,\alpha$ line with an atmosphere structure based on the model F99-C. This value is shown to be in a good agreement with observations  \citep[]{Tagirov2017}. Sunspots as well as faculae both contribute positively to variations of $Ly\,\alpha$, with the latter dominating because of the magnitude of their filling factors. The $Ly\,\alpha$ intensity of the model F99-P calculated with NESSY is also consistent with an average value emitted by active regions. Nevertheless, by using the filling factors reported above, CHRONOS still would underestimate the variability of the $Ly\,\alpha$ flux. The reason is that the filling factors are photospheric filling factors, whereas the $Ly\,\alpha$ line forms  in the transition zone typically above the height of 1000\,km. As the magnetic pressure within active regions decreases with height less steeply than the atmospheric pressure of the quiet Sun, the magnetic areas expand horizontally and the area covered by active regions increases with height. Therefore, to calculate realistic variations resulting from the variable solar activity, it is necessary to account for the larger areas of the active regions for spectral lines that emerge from layers above the photosphere. \citet{Schoell2011} estimated an average active area expansion factor of 2.4 by comparing SUMER images with MDI images. He also analysed three dimensional potential fields provided by  \citet{McIntosh2001} and computed an area expansion factor as a function of height, which monotonically increases from 1 at the phototosphere to 3.8 at a height of 1600\,km.  We derive an expansion factor of 3.0 to get agreement of the CHRONOS-reconstructed $Ly\,\alpha$ flux with satellite observations. This empirically adjusted factor is consistent with the investigations of \citet{Schoell2011}.
\section{Verification of total and spectral solar irradiance variability on decadal scale}
\begin{figure}
   \centering
   \includegraphics[width=9cm]{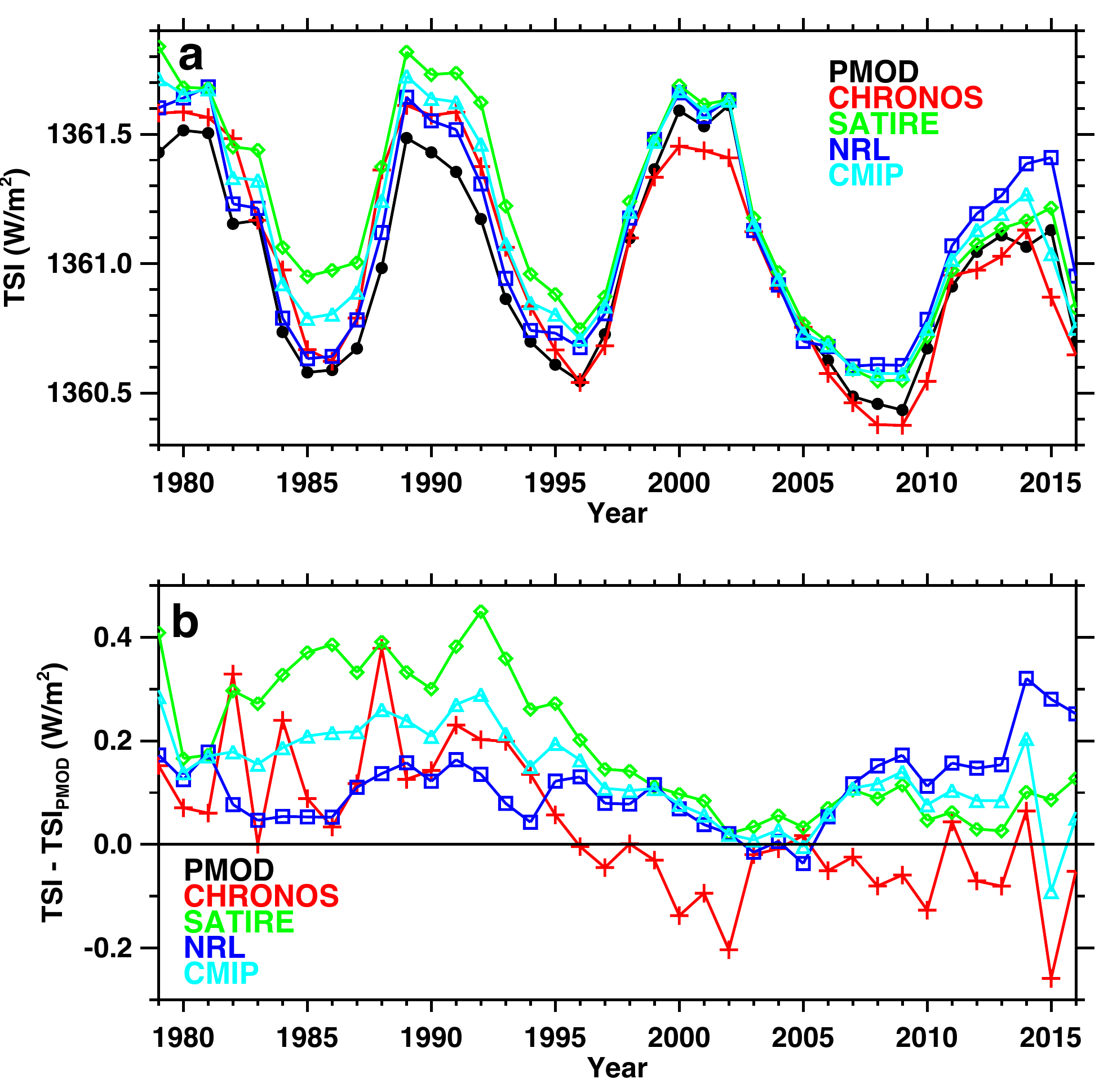}
   \caption{a) Evolution of the total solar irradiance (TSI, W/m$^2$) calculated with CHRONOS using the McCracken and Beer (2017, personal communication) solar modulation potential (red line, pluses) in comparison with the PMOD composite (black line, filed circles), the SATIRE-S (green line, open diamonds), the NRLSSI2 (blue line, open squares) models and solar irradiance prepared in the framework of the CMIP-6 project (light-blue line, open triangles). b) Deviations of the TSI reconstructions from the PMOD composite.}
   \label{Fig5}
   \end{figure}

The variations of solar irradiance due to the 11-year activity cycle are reproduced successfully by dedicated models based on active area filling factors based on full disc magnetograms and continuum images \citep{Ermolli2013}. However, for a long term reconstruction this high-quality modelling cannot be done because the necessary observations are not available for the pre-instrument era. Nevertheless, for many applications the 11-year variability is an important component and needs to be well-reproduced; despite the simplicity of our approach (Section\,2), one of our goals has been to reproduce the solar cycle as realistically as possible. To verify the performance of CHRONOS on this time scale, we investigate the total and spectral solar irradiance calculated for the satellite era (since 1978) and make comparisons with available observations and other model results. In Figure \ref{Fig5}  we compare the annual mean TSI calculated by CHRONOS based on the McCracken and Beer (2017, personal communication) solar modulation potential with the PMOD composite that combines satellite observations into a continuous record \citep{Froehlich2006}, and with the SATIRE-S \citep{Yeo2014} and NRLSSI2 \citep{Coddington2016} models as well as with the TSI data prepared within the framework of CMIP-6 project \citep{Matthes2016} and which is also based upon models. The version 42-65-1709 of the PMOD TSI composite was downloaded from ftp://ftp.pmodwrc.ch/pub/data/irradiance/composite/. The SATIRE (we use version S for satellite era and version T for the period from 1610 to present day) data source is http://www2.mps.mpg.de/projects/sun-climate/data. The NRLSSI2  dataset was acquired from ftp://data.ncdc.noaa.gov/cdr/solar-irradiance. CMIP-6 data are available from  (http://solarisheppa.geomar.de/cmip6).

The upper panel of Figure \ref{Fig5} represents the absolute annual mean TSI values while the lower panel shows the absolute difference between all data sets and the PMOD composite. The TSI calculated with CHRONOS is in good agreement with the PMOD composite during the last two solar minimum periods, while the cycle 23 maximum in 2000-2002 is underestimated by 0.1-0.2 W/m$^2$. The disagreement during the ascending and descending phases is generally smaller and does not exceed 0.2 W/m$^2$. In the early part of the composite (1979-1993) CHRONOS overestimates the PMOD TSI-composite up to 0.3 W/m$^2$. The comparison with the SATIRE and NRLSSI2 reconstruction models demonstrates that the CHRONOS deviation from the observations, despite its rather simplified treatment of the active regions, is similar to other models. The simplification of the active region treatment is reflected, however, in the behavior of the TSI deviation from the PMOD composite, which is less smooth than the other models. The absolute TSI levels of the other models are slightly higher (by up to 0.4 W/m$^2$) than the PMOD composite, but they reproduce the structure of the time dependence with better precision. This can be explained by the use of annual mean sunspot number in CHRONOS as the input for the solar irradiance calculations. Overall, we verify that the calculated TSI is in good agreement with models that are more sophisticated than CHRONOS and within 0.02\, \% relative to the PMOD composite, which is of the order of the uncertainty of the PMOD TSI-composite.

A correct simulation of TSI, which is the main component of the Earth's surface radiation balance, is necessary but not sufficient to characterize the solar irradiance forcing of the climate system. In the stratosphere the solar UV irradiance variability plays a crucial role in modulating the ozone mixing ratio and temperature via perturbations of photolysis and heating rates. 
For climate models the most important spectral intervals are the band 250-360\,nm, which controls stratospheric temperature \citep[e.g.,][]{Sukhodolov2014}, and the band 180-300\,nm, which is responsible for stratospheric ozone balance \citep[e.g.,][]{Ball2016a}. Therefore, a comparison of the SSI time evolution in two wavelength intervals (180-250 and 250-300\,nm) is a good indicator of the model performance.

\begin{figure}
   \centering
   \includegraphics[width=9cm]{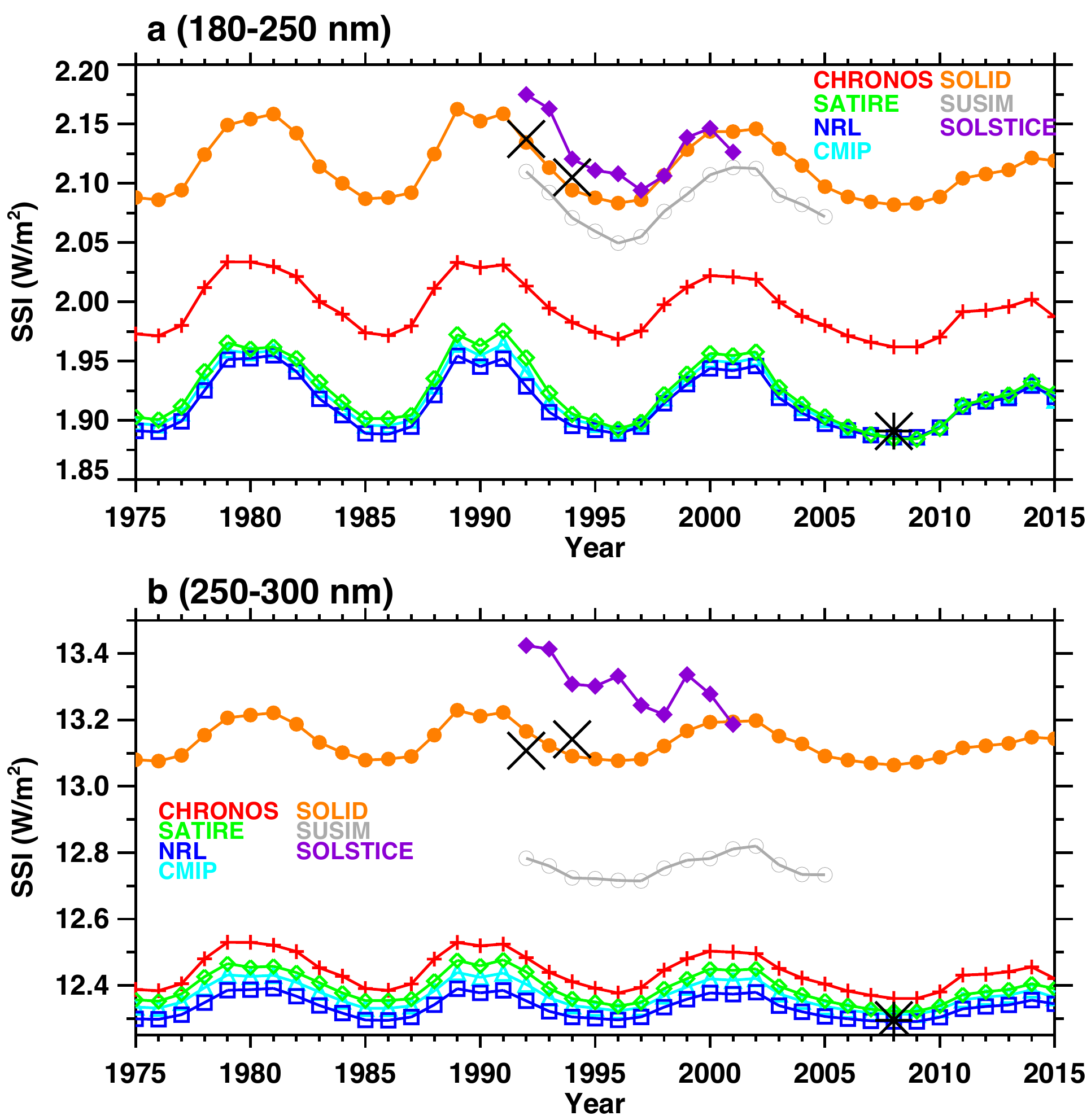}
   \caption{Evolution of the band-integrated annual mean spectral solar irradiance (W/m$^2$) calculated using CHRONOS (red line, pluses) in comparison with SATIRE-S (green line, open diamonds), NRLSSI2 (blue line, open squares), CMIP-6 (light-blue line, open triangles), UARS SUSIM (grey line, open circles), UARS SOLSTICE (violet line, filled diamonds), and SOLID composite (orange line, filled circles). The data from the reference spectral measurements in 1992, 1994 (ATLAS missions) and 2008 (WHI mission) are shown by crosses and the asterisk. Panels a and b show the results for the wavelengths intervals 180-250\,nm and 250-300\,nm respectively.}
   \label{Fig6}
   \end{figure}

For the observational verification of the CHRONOS reconstruction we use data measured by the SUSIM \citep{Rottman1993} and SOLSTICE \citep{Snow2005} instruments onboard the UARS satellite. Composites of observational data are compiled in the framework of the European SOLID project \citep{Schoell2016, Haberreiter2017}, and reference spectral measurements during ATLAS and WHI campaigns \citep{Thuillier2004,Woods2009} have been taken from  http://projects.pmodwrc.ch/solid/index.php/10-news-archive/36-ssi-datasets. Figure \ref{Fig6}  illustrates the temporal evolution of SSI after 1975, integrated over these bands. The quality of SSI data can be characterized by the accuracy in its representation of the absolute values and in the reproduction of the variability during the solar activity cycles. Figure \ref{Fig6} shows substantial disagreement in the absolute level of up to 15\,\% between the models and observations, which may lead to potential errors in the calculation of the thermal and chemical states in the terrestrial stratosphere. In particular, while solar irradiance in both wavelength bands from SATIRE, NRLSSI2, CMIP-6 and WHI data agree well in their absolute values, they are much lower compared to other observations. SSI calculated by the CHRONOS model in the 180-250\,nm region lies between two different groups of data and is close to other models and the WHI value for the 250-300\,nm spectral interval. The amplitude of the solar cycle is in good agreement for all models and observations. As in the TSI case, CHRONOS does not resolve the fine structure of the SSI temporal evolution. In particular, the observed decrease at solar maximum in 1990 is not well reproduced. 

\begin{figure}
   \centering
   \includegraphics[width=9cm]{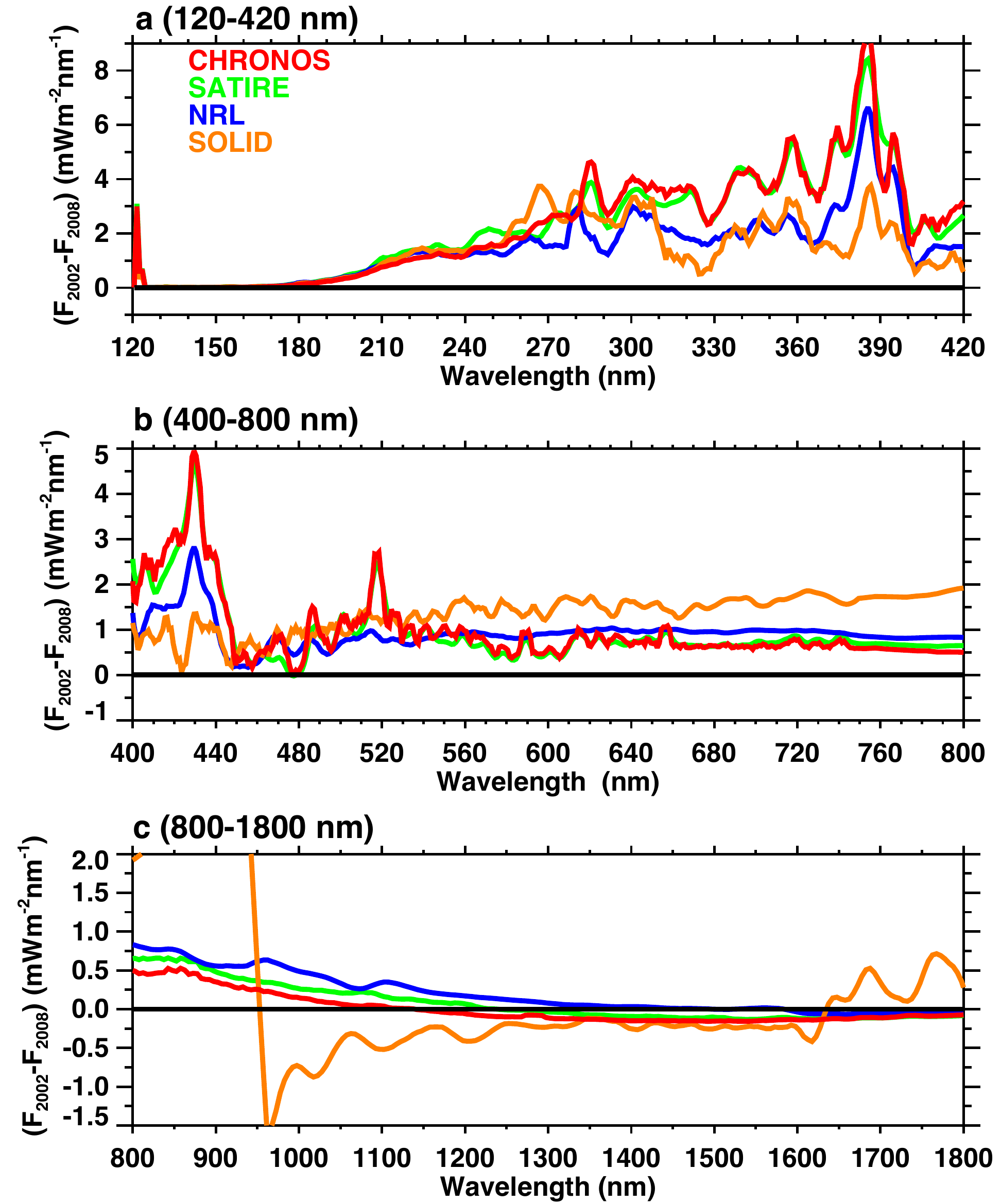}
   \caption{Spectral solar irradiance (mW/m$^2$/nm) difference between years 2002 and 2008 calculated with CHRONOS using PHI-MC17 solar modulation potential (red line) in comparison with the results of SATIRE-S (green line) and NRLSSI2 (blue line) models and SOLID composite (orange line). The original time series were smoothed using the 5\,nm window for better visibility.}
   \label{Fig7}
   \end{figure}

Figure \ref{Fig7} illustrates the absolute changes of SSI between 2002 (cycle 23 maximum) and 2008 (cycle minimum) for three wavelength intervals covering the range 120-1800\,nm, which characterises the irradiance variability during solar activity cycles. The variability at longer wavelengths is small and not shown here. The SSI variability simulated with the CHRONOS model is very close to SATIRE-S estimates in these spectral ranges. Both physics-based models are in remarkable agreement in reproducing the well-known main features of SSI variability in the UV range (120-420\,nm) such as enhanced variability from 270 to 400\,nm with the maximum around the violet system of the CN molecule at $\sim$390\,nm \citep{Shapiro2015}. At the same time, NRLSSI2 and SOLID underestimate UV variability between 300 and 400 nm compared to SATIRE-S. The reason for the lower variability in the NRLSSI2 reconstruction was explained by \citep{yeo2015} and relates to not considering uncertainties in the applied regression model used to build the irradiance model. The SOLID composite disagrees with all models above 250 nm since its quality is affected by the problems with SORCE data during the 2003-2008 period \citep{Haberreiter2017}. In the visible part of the solar spectrum good overall agreement between models is seen, in contrast to substantially higher variability of the SOLID composite. The NRLSSI2 model yields variability for wavelengths longer than 530\,nm that is higher than the other models; further the NRLSSI2 model does not reproduce the emission band around 490\,nm. In the infrared part of the solar spectrum all models agree well but the SOLID composite has a transition to negative variability here.

\section{The long-term evolution of solar irradiance}

\begin{figure}
   \centering
   \includegraphics[width=9cm]{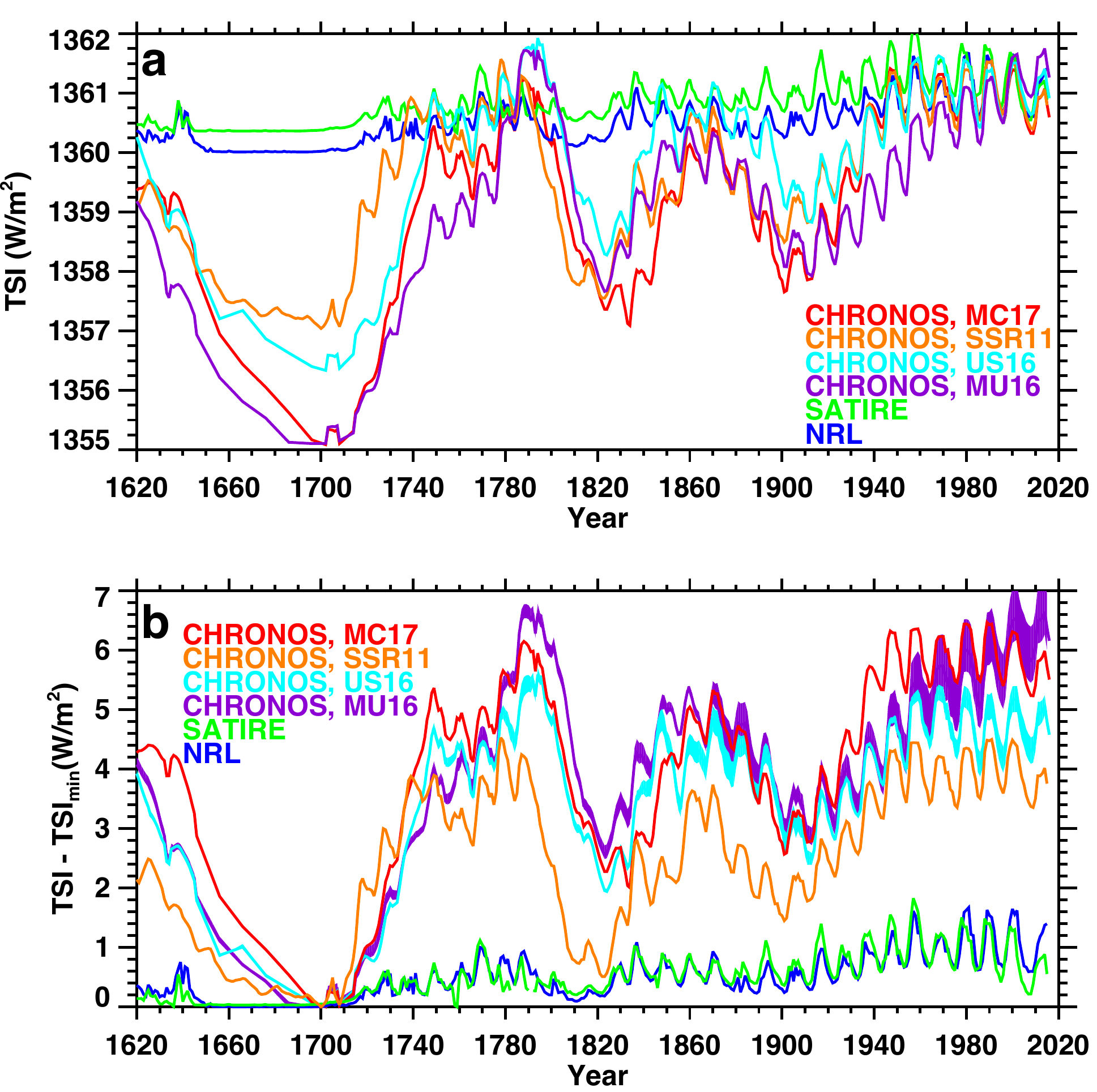}
   \caption{a) Temporal evolution of the total solar irradiance (TSI,\,W/m$^2$) calculated with CHRONOS using solar modulation potentials from SSR11 (orange line), PHI-MC17 (red line), PHI-US16 (light blue line) PHI-MU16 (violet line) in comparison with the SATIRE-T (green line) and NRLSSI2 (blue line) model outputs. b) Deviation of TSI (W/m$^2$) from the minimal values for the same models. The shading around blue and violet lines represent uncertainty range in the solar modulation potential reconstructions.  }
   \label{Fig8}
   \end{figure}

The TSI time series reconstructed by CHRONOS, SATIRE and NRLSSI2 models are shown in Figure \ref{Fig8}a in absolute values and in Figure \ref{Fig8}b as the deviation from minimal values during the considered period 1620-2015. The CHRONOS results are represented by four curves calculated using the four solar modulation potential datasets introduced in Section 2.1. All four reconstructions have an overall agreement in respect to periods of lower solar activity during Maunder, Dalton, and Gleisberg minima as well as the period of high and persistent TSI from 1940 until 2002 \citep[the Modern grand maximum, see][]{Solanki2004}. 

The TSI increase from Maunder minimum to 2008 in the CHRONOS models is much higher than for the NRLSSI2 and SATIRE results for all solar modulation potential versions because of the different treatment for the evolution of the quiet Sun irradiance. The amplitude of the CHRONOS reconstruction reaches almost 5\,W/m$^2$, which is lower than the 6\,W/m$^2$ suggested by SSR11 as a mean estimate. The change of the TSI variability amplitude in respect to the version of SSR11 is explained by the use of the slightly warmer atmospheric structure model B as described in Section 2.1, in combination with the use of new solar modulation potentials. Figure \ref{Fig8}b illustrates the comparison between the different CHRONOS versions and other reconstructions. Keeping the solar modulation potential as used by SSR11 illustrates the change from model A to model B (orange line), which reduces the long-term variability amplitude from 6\,W/m$^2$ to 3.4\,W/m$^2$. On the other hand, the use of the PHI-MC17 solar modulation potential re-enhances the amplitude by about 1.5\,W/m$^2$. As a consequence of the assumption that the quiet Sun irradiance is modulated in synchronisation with the heliospheric modulation potential, there is a clearly visible TSI variation during the Maunder minimum. While the NRLSSI2 and SATIRE models reconstruct a constant flat TSI during the Maunder minimum, the CHRONOS model yields a strong variation, reflecting the fact that, even though sunspots virtually disappeared, heliospheric activity was still varying during the Maunder minimum \citep{Beer1998, Owens2012}. 

The application of three solar modulation potential time series (see Sect.2.1) for the reconstruction of the TSI evolution allows for, in part, an estimation of uncertainty in the computed solar forcing formed from the dispersion between reconstructions. However, due to several additional assumptions in the applied reconstruction, the global uncertainty could be larger than estimated. The TSI changes calculated with the PHI-US16 and PHI-MU16 data agree well during the Maunder minimum onset and reveal about 1.0\,W/m$^2$ weaker TSI decline in comparison to PHI-MC17. The TSI obtained with PHI-US16 exceeds the TSI calculated with PHI-M16 and PHI-MC17 during the Maunder minimum because the deviation of the solar modulation potential from its minimum value in 1466 is higher for this data set. During the recovery phase (1700-1780 period) the PHI-MC17 and PHI-US16 data give a local maximum around 1750, which is in contrast with the reconstruction using PHI-MU16, for which TSI steadily increases. In 1780, however, TSI reconstructed from all three data sets converge reaching of the order of 5-6\,W/m$^2$ difference to the minimum value. Similar partial agreement and disagreement remains up to 1900. In the first half of the 20$^{th}$ century all reconstructions yield a similar pace of TSI increase. After 1950, the TSI from PHI-MC17 remains high and starts to decrease after 1980, while TSI from PHI-US16 stays almost constant after 1940 and TSI from PHI-MU16 keeps rising until 2000. The TSI difference between the Maunder minimum value to the solar cycle minimum in 2009 varies from 3.8 to 6.2\,W/m$^2$ and this interval can be taken as an uncertainty range. The comparison of all CHRONOS data with SATIRE and NRLSSI2 demonstrates that the choice of the particular solar modulation potential data set plays a secondary role compared to the effect of the choice of the minimum state of the Sun for the long-term evolution of the quiet Sun irradiance.

\begin{figure}
   \centering
   \includegraphics[width=9cm]{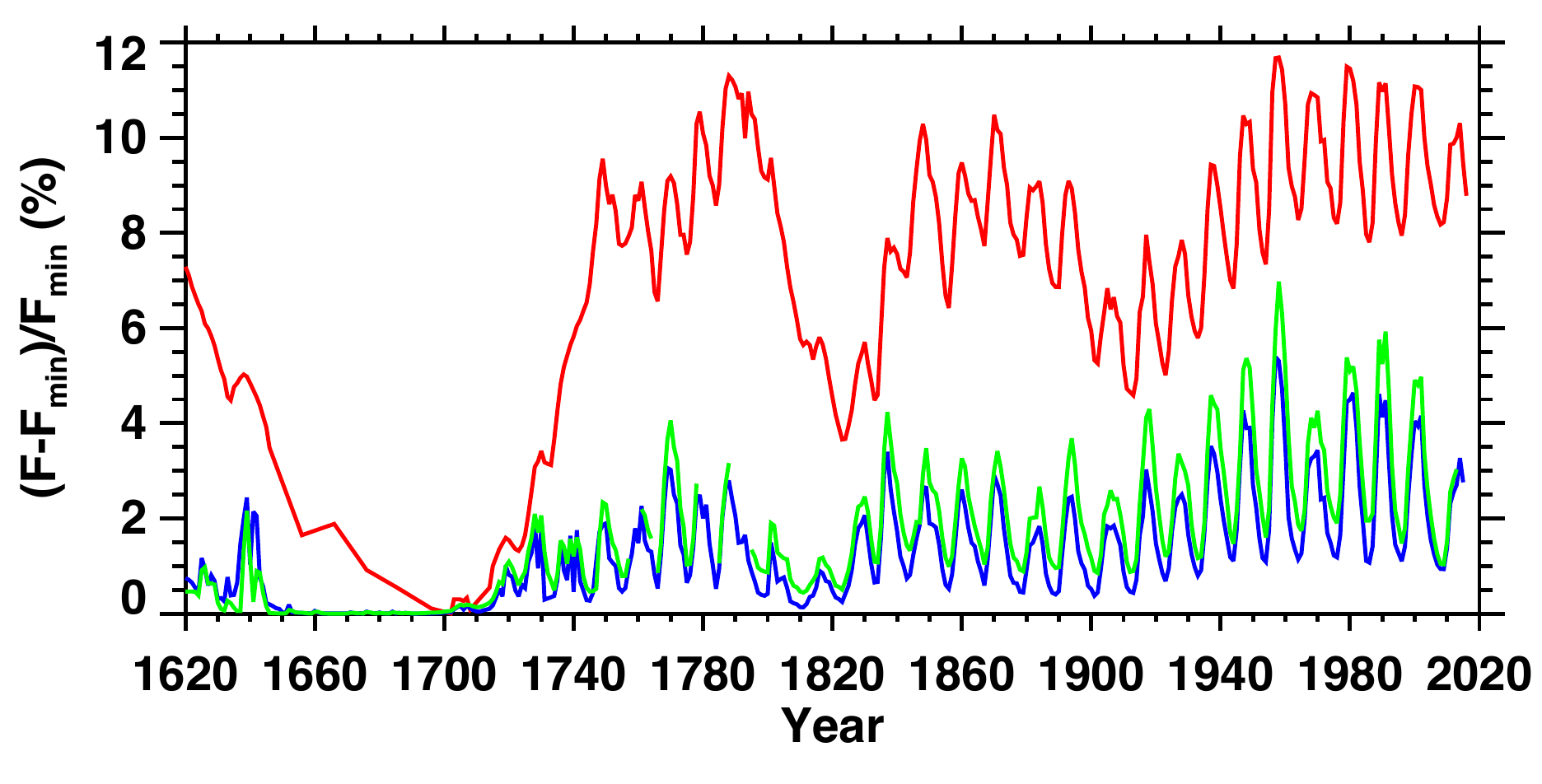}
   \caption{Relative deviation (in \%) of SSI integrated over the 180-250\,nm spectral interval from its minimum value as calculated with CHRONOS using PHI-MC17 solar modulation potential (red line) in comparison with those by SATIRE-T (green line) and NRLSSI2 (blue line) models.}
   \label{Fig9}
   \end{figure}

Similar comments as above applied to the SSI reconstruction. Figure \ref{Fig9} shows that in the spectral interval 180-250\,nm the CHRONOS model yields an increase of $\sim$9\,\% from Maunder minimum to the solar cycle minimum in 2009, while the increase suggested by SATIRE and NRLSSI2 models does not exceed 1\,\%. For the spectral interval 250-300\,nm the variability is the same as for the 180-250\,nm interval but with half the magnitude of the variations (not shown). 

\begin{figure}
   \centering
   \includegraphics[width=9cm]{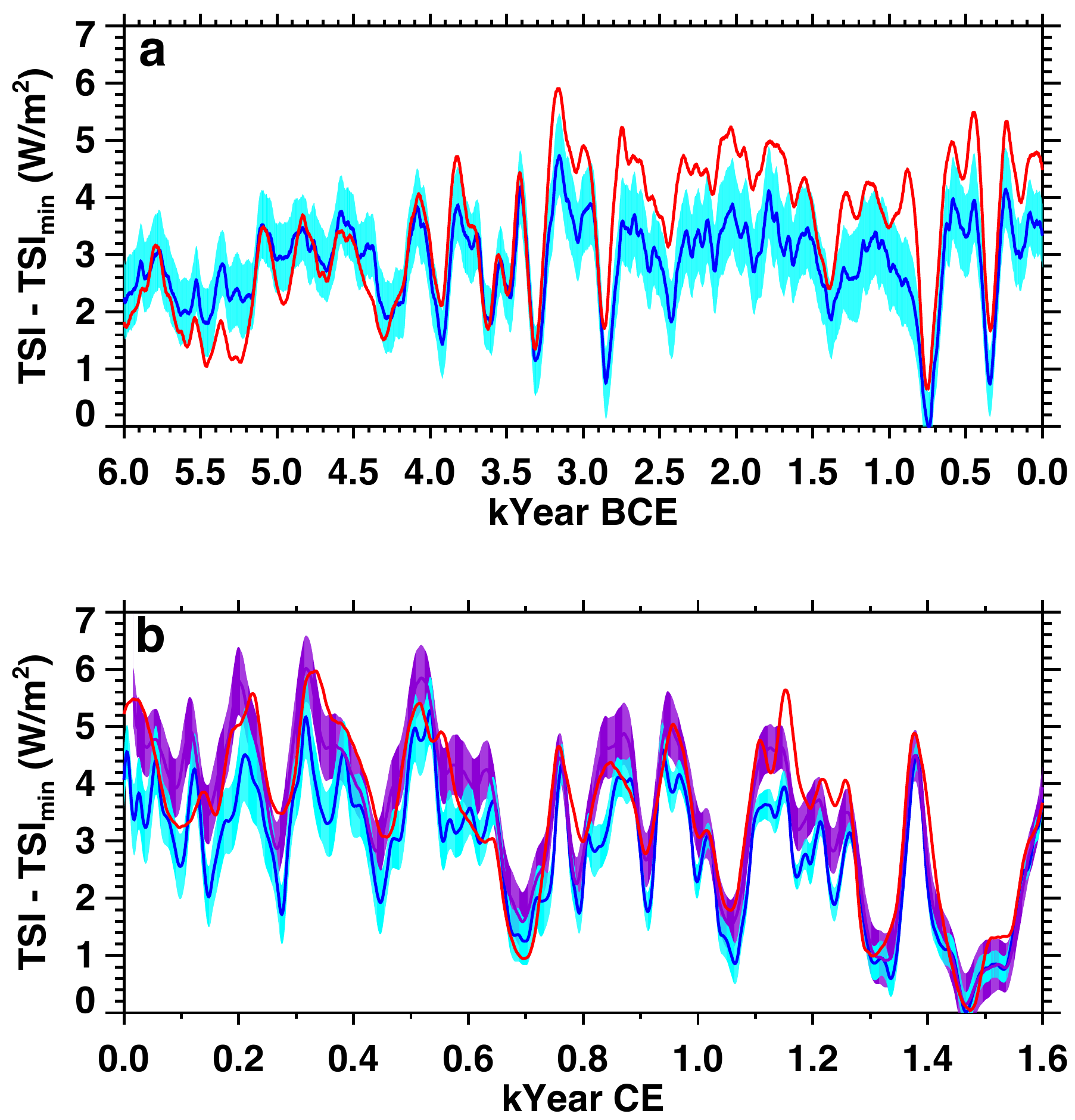}
   \caption{Deviation of TSI (W/m$^2$) from its minimal value calculated with CHRONOS using PHI-MC17 (red), PHI-US16 (blue) and PHI-MU16 (violet) versions of the solar modulation potential. The time series were smoothed using 100 or 11 years window for the upper and lower panels, respectively. The shading around blue and violet lines represent uncertainty range in the solar modulation potential reconstructions. }
   \label{Fig10}
   \end{figure}

Figure \ref{Fig10} illustrates the TSI variability on millennial time scales calculated with the CHRONOS model using all three above-mentioned solar modulation potential time series. The BCE period is covered by PHI-MC17 and PHI-US16 datasets. The two reconstructed TSI time series exhibit large (up to 6.1\,W/m$^2$) variability and mostly agree within the uncertainty range, however PHI-MC17 gives a lower TSI value before 4.2 kYear BCE and higher values after 3 kYear BCE. This may be related to the different archeo/paleomagnetic models used in these reconstructions of the solar modulation potential. We note that the largest local TSI spikes/dips, corresponding to grand solar maxima/minima, is also seen. However, the agreement is not good for two periods around 5.2 kYear BCE and 3.5-0.9 kYear BCE when PHI-US16 gives higher/lower TSI values. 

The CE period is covered by all three solar modulation potential datasets. According to all three reconstructions, TSI reaches its minimum value around 1450 (Sp\"{o}rer minimum), while the maximum value (up to 6.5\,W/m$^2$) appears between 200 and 600 CE. After that, all data sets show a gradual decrease in the mean TSI level until 1600 CE. We again note that the largest local TSI spikes is also agree well. Some minor discrepancies between the TSI series are related to the difference in the modulation potential series used here.  

The temporal evolution of SSI is similar to that of TSI because they both are modelled using the secular variability of the solar modulation potential, but the magnitude of the relative changes maximises in the short wavelength, reaching more than 50\,\% in the Lyman-$\alpha$ line and decreases with wavelength to less then 1\,\% in the infrared part (not shown).  

\section{Conclusions}
In this paper we present a new reconstruction of TSI and SSI covering the period from 6000 BCE to 2015 CE using the {\em Code for the High ResolutiOn recoNstructiOn of Solar spectral irradiance} (CHRONOS).The model retrieves the filling factors for bright and dark features on the solar disk, using sunspot number, and combines the irradiance output of spectral synthesis code NESSY, which calculates the spectral solar irradiance for a set of given solar atmosphere models, to produce a time-varying reconstruction of solar irradiance. The new reconstruction, compared to SSR11, shows improved solar irradiance variability on decadal time scales, which facilitates a better agreement with direct observations during the satellite era. We demonstrate that the performance of CHRONOS during the satellite period is comparable on annual time-scale to other solar irradiance reconstructions over the recent, observed period that use more sophisticated treatment of the solar active regions. Using the updated model, and new proxy data for the solar modulation potential, we suggest a multi-centennial solar forcing whose magnitude is 25-40\,\% smaller than that proposed by SSR11, but still significantly higher than the results of the other groups. This change in the magnitude of the irradiance variation with respect to SSR11 is a result of applying a new solar structure model for the minimum state of the quiet Sun and a revision solar modulation potential data. In agreement with the assessment of \citet{Judge2012} the substitution of F99 model A used in SSR11 for model B reduces the amplitude of the solar forcing from SSR11 by about a factor of two. This correction is partly compensated by the use of three independently reconstructed solar modulation potential data sets, which enhance the TSI increase from the Maunder minimum to the solar cycle minimum in 2009 to 3.8-6.2\,W/m$^2$, depending on the applied solar modulation potential time series. The SSI increase during the same period strongly depends on the wavelength and reaches 9\,\% for SSI integrated over 180-250\,nm. The CHRONOS reconstruction substantially exceeds the magnitudes of the TSI and SSI increase reconstructed by SATIRE and NRLSSI2 models on centennial time scales (0.2-0.6\, W/m$^2$ and 1\,\%). Observational proof of which reconstruction might be more realistic will eventually come from future observations as the Sun's irradiance evolves. If the underlying physically relationships of the CRONOS model is correct, the solar radiance will decrease proportionally to the solar modulation potential as described in Section 2.1. As it is anticipated that in the next minimum there will be a decrease of the 22-year mean level of the solar modulation potential compared to the previous minimum, there may be an opportunity to measure and test the difference.

On multi-millennial time scales, the TSI time series reconstructed using different solar modulation potentials are generally in a good agreement except for some limited periods (e.g., 3200-0 BCE). They unanimously point to the lowest TSI value occurring around 1450 CE, and the highest TSI (exceeding 6.0\,W/m$^2$) appears between 200 and 500 CE. Overall, the magnitude of the TSI and SSI variability on the millenial time scale is similar or slightly higher than during past 400 years of CE.

The uncertainty of our TSI and SSI reconstructions is substantial. The use of different solar modulation potential data sets yields an uncertainty of almost a factor of two as illustrated in Fig.\ref{Fig8}. In addition, a similar magnitude of the uncertainty can be attributed to the choice of the solar atmosphere model PW-B to represent the minimal activity state of the Sun. The range defined by the three reconstruction versions can be seen as representing the uncertainty estimate. The PHI-US16 yields the smallest forcing since the Maunder minimum of about 4\,W/m$^2$ (see Fig.\ref{Fig8}b) and if we account for the uncertainty of the minimum model, this could reduce to 3\,W/m$^2$. The reconstruction based on PHI-MU16 yields the largest TSI increase of 6\,W/m$^2$, which could be as large as 7\,W/m$^2$ if a solar atmosphere model that gives a value at the lower end of the brightness distribution was selected, as shown in Fig.\ref{Fig1} for the solar minimum state.

The reconstructed SSI data sets, corresponding to the three solar modulation potentials that have been used as input to the CHRONOS, are available from the corresponding author by request.

\begin{acknowledgements}
The work was performed in the framework of the Swiss National Science Foundation project under grant agreement  CRSII2-147659 (FUPSOL II). ER was partly supported by SNSF projects 149182 (SILA) and 163206 (SIMA). AS acknowledges funding from the People Programme (Marie Curie Actions) of the European Union$'$s Seventh Framework Programme (FP7/2007-2013) under grant agreement No. 624817 and the European Research Council under the European Union$'$s Horizon 2020 research and innovation programme (grant agreement No. 715947). IU acknowledges ReSoLVE Centre of Excellence of the Academy of Finland (Project 272157).  RVT acknowledges funding from SNF grant 200020-153301.This work has benefited from the ROSMIC WG1 activity within the SCOSTEP VarSITI program. We would like to thank R. Muscheler and K. McCracken for providing their solar modulation potential reconstructions. We acknowledge fruitful discussions in the framework of the International Team 373 "Towards a Unified Solar Forcing Input to Climate Studies" supported by the International Space Science Institute (ISSI). We would like to thank N.Krivova for providing SATIRE fillings factors and for useful comments to earlier version of the manuscript. We would like to thank the anonymous referee for valuable comments/corrections that greatly improved the manuscript. We are also gratefull to Dr. W. Ball for language editing.

\end{acknowledgements}
%
%

   \bibliographystyle{aa} 

\end{document}